\documentclass[%
onecolumn,aip,jmp,amsmath,amssymb,eqsecnum,
reprint,%
]{revtex4-1}

\usepackage{graphicx}% Include figure files
\usepackage{dcolumn}% Align table columns on decimal point
\usepackage{bm}% bold math
\begin{document}

%\preprint{AIP/123-QED}

\title[Path Integral and Spectral Representations for Supersymmetric Dirac-Hamiltonians]{Path Integral and Spectral Representations for Supersymmetric Dirac-Hamiltonians}
%\thanks{Footnote to title of article.}

\author{G. Junker}
%\altaffiliation[Also at ]{Theorie 1Physics Department, XYZ University.}%Lines break automatically or can be forced with \\
\email{gjunker@eso.org}
\affiliation{European Organization for Astronomical Research in the Southern Hemisphere,\\
Karl-Schwarzschild-Strasse 2, D-85748 Garching, Germany%\\This line break forced with \textbackslash\textbackslash
}%

\author{A. Inomata}
\affiliation{%
Department of Physics, State University of New York at Albany,\\
Albany, NY 12222, USA%\\This line break forced% with \\
}%

\date{\today}% It is always \today, today,

\begin{abstract}
  The resolvent of supersymmetric Dirac Hamiltonian is studied in detail. Due to supersymmetry the squared Dirac Hamiltonian becomes block-diagonal whose elements are in essence non-relativistic Schr\"odinger-type Hamiltonians. This enables us to find a Feynman-type path-integral representation of the resulting Green's functions. In addition, we are also able to express the spectral properties of the supersymmetric Dirac Hamiltonian in terms of those of the non-relativistic Schr\"odinger Hamiltonians. The methods are explicitly applied to the free Dirac Hamiltonian, the so-called Dirac oscillator and a generalization of it. The general approach is applicable to systems with good and broken supersymmetry.
\end{abstract}

%Use showkeys class option if keyword
\maketitle

\newcommand{\rmi}{\textrm{i}}
\newcommand{\rmd}{\textrm{d}}
\newcommand{\balpha}{\bm{\alpha}}
\newcommand{\bsigma}{\bm{\sigma}}
\newcommand{\rme}{\textrm{e}}
\newcommand{\JPA}{J.\ Phys.\ A}
\newcommand{\RMP}{Rev.\ Mod.\ Phys.\ }
\newcommand{\PRL}{Phys.\ Rev.\ Lett.\ }
\newcommand{\JMP}{J.\ Math.\ Phys.\ }

\section{Introduction}
The idea of supersymmetry (SUSY) in non-relativistic quantum mechanics was first introduced by Nicolai \cite{Nic76} in 1976 for analyzing spin systems with a $(0+1)$-dimensional version of supersymmetric quantum field theories. SUSY quantum mechanics became popular when Witten's model was introduced in 1981 \cite{Wit81}. Since then SUSY has become an important tool in studying properties of models in quantum mechanics and statistical physics \cite{Junker1996}. Jackiw \cite{Jackiw1984} observed that the square of a 2-dimensional Dirac particle subjected to an external magnetic field is directly related to the non-relativistic Pauli-Hamiltonian of the same system. This observation triggered a more extended study of SUSY for the Dirac Hamiltonians \cite{Hughesetal86,Cooper1988}. Indeed, under certain conditions the Dirac Hamiltonian may be treated as supercharge of some non-relativistic system in SUSY quantum mechanics\cite{Thaller1988,Beckers1990,Thaller1991}. Such a SUSY structure in turn proofs to be very useful in studying the spectral properties of the Dirac Hamiltonian. For an overview of relativistic quantum systems exhibiting SUSY, see ref.\ [\onlinecite{TH1992}].

In recent years, SUSY has also become a key concept in characterizing condensed matter systems known as topological superconductors\cite{Groverelal2014} and also plays a crucial role in explaining quantum phenomena in carbon-based nano-structures like nano-tubes and graphene. In the latter the two-dimensional Dirac Hamiltonian is used in characterizing its electronic transport properties.\cite{Sarma2011} Here SUSY plays a crucial role, for example, in understanding phenomena like the unconventional quantum Hall effect.\cite{Ezawa2006}

In the present paper, we study the resolvent of the Dirac Hamiltonian with the SUSY structure. We show that the Green's functions resulting from the resolvent can be determined by means of path integrals for the effective non-relativistic Lagrangians. The spectral properties of the Dirac Hamiltonian can be derived in a simple way from those of the associated Schr\"odinger Hamiltonians. We also present explicit applications to a few examples of the Dirac system.

In Section II, a stage is set for studying the SUSY properties of the Dirac Hamiltonian. In particular the spectral properties of a supersymmetric Dirac Hamiltonian are shown to be given explicitly via those of a corresponding non-relativistic $N=2$ SUSY Hamiltonian.
Section III deals with the path integral representation of the Green's functions based on the effective Schr\"odinger Hamiltonian. The path integral used is formally equivalent to Feynman's, in which the classical action is replaced by the effective action corresponding to the effective Schr\"odinger Hamiltonian. In Sections IV to VI are devoted to solving examples which include the free Dirac electron, the Dirac oscillator and a generalized Dirac oscillator. Concluding remarks are given in Section VII.

\section{Supersymmetric Dirac Hamiltonians}
Supersymmetric (SUSY) quantum mechanics, as defined by Witten \cite{Wit81}, is characterized by Hamiltonian $H_{SUSY}$ and a set of self-adjoint operators $Q_i$ ($i=1,2,\ldots, N$) acting on some Hilbert space ${\cal H}$ and satisfying (see, e.g., ref.\ [\onlinecite{Junker1996}]),
\begin{equation}\label{SUSYAlgebra}
  \{ Q_i, Q_j \} = H_{SUSY} \delta_{ij}, \quad i,j=1,2,\ldots,N \, ,
\end{equation}
where $\{A,B\}:=AB+BA$ denotes the anti-commutator.

To study the SUSY-structure of the  Dirac system, we restrict ourselves to the case of $N=2$ which involves only the three operators, $Q_1,Q_2$ and $H_{SUSY}$ acting on ${\cal H}$. Here, instead of dealing with $Q_1$ and $Q_2$, we define complex supercharges
\begin{equation}\label{complexQ}
  Q = \frac{1}{\sqrt{2}}\left(Q_1 + \rmi Q_2\right)\, , \quad  Q^\dag = \frac{1}{\sqrt{2}}\left(Q_1 - \rmi Q_2\right)\, .
\end{equation}
These operators and the SUSY Hamiltonian close the superalgebra \cite{Nic76},
\begin{equation}\label{SUSYAlgebra2}
  \{ Q, Q^\dag \} = H_{SUSY} \, , \quad Q^2 = 0 =\left(Q^\dag\right)^2 \, .
\end{equation}

In addition, we introduce the so-called Witten operator $W$ which is a unitary (non-trivial) involution on ${\cal H}$, commutes with the SUSY Hamiltonian and anticommutes with the complex supercharges, that is,
\begin{equation}\label{SUSYAlgebra3}
\begin{array}{c}
  W^\dag = W \, , \quad W^2=1 \, , \quad [W,H_{SUSY}]=0 \, , \\[2mm]
  \{W,Q\} = 0 =\{W,Q^\dag\} .
\end{array}
\end{equation}
See ref.\ [\onlinecite{Junker1996}]. The projection operators defined by $P^\pm=(1\pm W)/2$ decompose ${\cal H}$ into the eigenspaces ${\cal H}^\pm$ of the Witten operator with eigenvalues $\pm 1$. Namely,
\begin{eqnarray}
  {\cal H}    &=& {\cal H}^+\oplus {\cal H}^-\, ,\\
  {\cal H}^\pm &=& P^\pm {\cal H}= \{|\Psi^\pm\rangle\in{\cal H}:W|\Psi^\pm\rangle=\pm|\Psi^\pm\rangle\} .
\end{eqnarray}

In the representation in which $W$ is diagonal, i.e.,
\begin{equation}\label{W}
  W =
  \left(
  \begin{array}{cc}
    1 & 0 \\
    0 & -1
  \end{array}
  \right),
\end{equation}
the complex supercharges take the form,
\begin{equation}\label{Q}
  Q =
  \left(
  \begin{array}{cc}
    0 & D \\
    0 & 0
  \end{array}
  \right),\quad
  Q^\dag =
  \left(
  \begin{array}{cc}
    0 & 0 \\
    D^\dag & 0
  \end{array}
  \right),
\end{equation}
where $D$ is not self-adjoint in general. The SUSY Hamiltonian is usually given by
\begin{equation}\label{Hsusy}
  H_{SUSY}=\{Q,Q^\dag\}=
  \left(
  \begin{array}{cc}
    DD^\dag & 0 \\
    0& D^\dag D
  \end{array}
  \right).
\end{equation}
However, more generally, within the framework of the SUSY algebra (\ref{SUSYAlgebra3}), we may choose the SUSY Hamiltonian in the form,
\begin{equation}\label{Hsusy2}
  H_{SUSY}=a\{Q,Q^\dag\} + \varepsilon_0\textbf{1}=
  \left(
  \begin{array}{cc}
    aDD^\dag +\varepsilon_0& 0 \\
    0& aD^\dag D+\varepsilon_0
  \end{array}
  \right),
\end{equation}
where $a$ and $\varepsilon_0$ are arbitrary real constants with $a>0$. In studying the Dirac problem such a modification is not imperative but useful. We treat the two diagonal elements $H^+_{SUSY}=aDD^\dag+\varepsilon_0$ and $H^-_{SUSY}=aD^\dag D+\varepsilon_0$ as the SUSY partner Hamiltonians. As in the usual SUSY case the partner Hamiltonians are isospectral. Let $\varepsilon^\pm$ and $|\Psi_\varepsilon^\pm\rangle$ denote the eigenvalues and eigenstates of $H^\pm_{SUSY}$, that is,
\begin{equation}\label{SpecPropHsusy}
  H^\pm_{SUSY}|\Psi_\varepsilon^\pm\rangle = \varepsilon^\pm |\Psi_\varepsilon^\pm\rangle\,.
\end{equation}
It is obvious that $\varepsilon^\pm\geq \varepsilon_0$ and $\varepsilon^+=\varepsilon^-=\varepsilon$ for $\varepsilon>\varepsilon_0$. The eigenstates $|\Psi_\varepsilon^\pm\rangle$ for the same eigenvalue $\varepsilon>\varepsilon_0$ are related by the SUSY transformations
\begin{equation}\label{SUSYtrafogen}
\begin{array}{rl}
  D^\dag|\Psi_\varepsilon^+\rangle = &\sqrt{(\varepsilon-\varepsilon_0)/a}|\Psi_\varepsilon^-\rangle\, ,\\[2mm]
  D|\Psi_\varepsilon^-\rangle = & \sqrt{(\varepsilon-\varepsilon_0)/a}|\Psi_\varepsilon^+\rangle \,.
\end{array}
\end{equation}
For $\varepsilon=\varepsilon_0$ there might exist eigenstates in ${\cal H}^-$ or in ${\cal H}^+$ or in both. In that case SUSY is said to be unbroken. If $\varepsilon_0$ does not belong to the spectrum of $H^-_{SUSY}$ and $H^+_{SUSY}$ SUSY is said to be broken.

Now we consider the Dirac Hamiltonian
\begin{equation}\label{HD}
  H_D=c\balpha\cdot\left(\mathbf{p}-\frac{e}{c}\mathbf{A}\right)+\beta m_0c^2.
\end{equation}
Here $\balpha$ and $\beta$ are the $4\times 4$ Dirac matrices, $c$ is the speed of light, $m_0$ the rest mass and $e$ the charge of an electron. Notice that the scalar potential is missing in (\ref{HD}). In the present paper, we study only the case where the scalar potential is absent.

We employ the standard representation of the Dirac matrices
\begin{equation}\label{alpha&beta}
  \balpha=\left(
  \begin{array}{cc}
    0 & \bsigma \\
    \bsigma & 0
  \end{array}
  \right)\, ,\quad
  \beta=\left(
  \begin{array}{cc}
    1 & 0 \\
    0 & -1
  \end{array}
  \right)\, ,
\end{equation}
where $\bsigma$ is the unit vector based on the Pauli matrices, and let
\begin{equation}\label{D}
  D=c\bsigma\cdot\left(\textbf{p}-\frac{e}{c}\textbf{A}\right)\, ,\quad M_0=m_0c^2\, .
\end{equation}
Then we can express the Dirac Hamiltonian (\ref{HD}) in the form
\begin{equation}\label{HD2}
  H_D=\left(Q+Q^\dag\right)+M_0W=\sqrt{2}Q_1+M_0W
\end{equation}
or
\begin{equation}\label{HD3}
  H_D=\left(
  \begin{array}{cc}
    M_0 & D \\
    D^\dag & -M_0
  \end{array}
  \right).
\end{equation}
The operator $D$ specified by (\ref{D}) is not self-adjoint unless the vector potential $\textbf{A}$ is real.

Evidently, in Witten's sense, $H=(H_D-M_0W)^2$ is the SUSY Hamiltonian of $N=1$ with supercharge $Q_1=(H_D-M_0W)/\sqrt{2}$. In the approach of Thaller \cite{Thaller1988} and Beckers and Debergh \cite{Beckers1990}, the Dirac Hamiltonian itself is supersymmetric in the sense that it has the odd part $H_D^{odd}=Q+Q^\dag$ and the even part $H_D^{even}=M_0W$ with respect to $W$, obeying $\{H_D^{odd},W\}=0$ and $[H_D^{even},W]=0$.
In the preceeding sections, we shall, however, treat the Dirac Hamiltonian (\ref{HD}) or (\ref{HD2}) in the framework of SUSY quantum mechanics of $N=2$. To this end, we pay attention to the squared Dirac operator which turns out to be (block) diagonal
\begin{equation}\label{HD^2}
\begin{array}{rl}
  \displaystyle
  H^2_D &= \{Q,Q^\dag\} + M_0^2 \\[2mm]
  \displaystyle &=
  \left(
  \begin{array}{cc}
    D D^\dag +M_0^2 & 0 \\
    0 & D^\dag D +M_0^2
  \end{array}
  \right).
\end{array}
\end{equation}
If $H_D^2$ is taken as the SUSY Hamiltonian, then its diagonal elements are SUSY partner.

Since $DD^\dag$ and $D^\dag D$ behave as $\sim c^2\textbf{p}^2$, we introduce a mass parameter $m>0$ to make the squared Dirac operator look like a Schr\"odinger Hamiltonian; namely
\begin{equation}\label{HS}
  H_{SUSY}=\frac{1}{2mc^2}H^2_D=
  \left(
  \begin{array}{cc}
    H^+_{SUSY} & 0 \\
    0 & H^-_{SUSY}
  \end{array}
  \right),
\end{equation}
where
\begin{equation}\label{HS+-}
\begin{array}{c}
   \displaystyle
   H_{SUSY}^+=\frac{1}{2mc^2}\left(DD^\dag+M^2_0\right)\,,\\[4mm]
   \displaystyle
   H_{SUSY}^-=\frac{1}{2mc^2}\left(D^\dag D+M^2_0\right)\,,
\end{array}
\end{equation}
acting on the subspaces ${\cal H}^\pm$, respectively. In other words, we choose $a=1/2mc^2$ and $\varepsilon_0=M^2_0/2mc^2$ in (\ref{Hsusy2}).
The diagonal elements of the squared Dirac operator modified with the mass parameter are indeed the Schr\"odinger Hamiltonians for a non-relativistic system of mass $m$ in form. The mass parameter $m$ introduced above may be identified with the electron rest mass $m_0$, but will be left unspecified for a while to take note of its formal and arbitrary nature.

Now we can utilize these Schr\"odinger operators as SUSY partner Hamiltonians by regarding $H_{SUSY}=H_D^2/2mc^2$ as the SUSY Hamiltonian. Furthermore, use of the Schr\"odinger Hamiltonians enables us to pursue the Feynman-type path integral representation for the Dirac systems as will be discusses in the proceeding sections.

It had been noted by Thaller \cite{Thaller1988}, see also ref.\ [\onlinecite{Beckers1990},\onlinecite{Thaller1991}], that the Dirac operator can also be diagonalized by a unitary  Foldy-Wouthuysen transformation $U_{\rm FW}$. Let
\begin{equation}\label{UFW}
  U_{\rm FW}= \frac{|H_D|+WH_D}{\sqrt{2H^2_D+2M_0|H_D|}}
\end{equation}
be a unitary operator, where
\begin{equation}\label{|HD|}
  |H_D|=\sqrt{H_D^2}=\sqrt{2mc^2H_{SUSY}}
\end{equation}
is an even operator, $[W,|H_D|]=0$. It is easy to show that
\begin{equation}\label{Hdiagonal}
\begin{array}{rl}
  H_{\rm FW}&=U_{\rm FW} H_DU_{\rm FW}^\dagger= W|H_D|\\[2mm]
  &\displaystyle =
  \left(
  \begin{array}{cc}
    \sqrt{D D^\dagger+M_0^2} & 0 \\
    0 & -\sqrt{D^\dagger D +M_0^2}
  \end{array}
  \right) ,
\end{array}
\end{equation}
which implies that the spectrum of $H_D$ can be obtained from the spectrum of $H_{\rm FW}$. To be a bid more explicit let us denote the upper and lower diagonal elements of $H_{\rm FW}$ by $H^\pm_{\rm FW}$. We may express them via the SUSY Hamiltonians $H^\pm_{SUSY}$ as follows
\begin{equation}\label{HFWpm}
  H^\pm_{\rm FW}=\pm\sqrt{2mc^2H^\pm_{SUSY}}.
\end{equation}
Hence the spectrum of $H_D$ which is identical to that of $H_{\rm FW}$ can in turn be obtained from that of $H_{ SUSY}$.
% as
%\begin{equation}\label{spectrumHD}
%\begin{array}{l}
%  \textrm{spec}(H_D) =\\
%  ~ =  \textrm{spec}\left(\sqrt{DD^\dag+M^2_0}\right)\cup\textrm{spec}\left(-\sqrt{D^\dag D+M^2_0}\right)\, . \\
%  ~ =  \textrm{spec}\left(\sqrt{2mc^2 H^+_{SUSY}}\right)\cup\textrm{spec}\left(-\sqrt{2mc^2 H^-_{SUSY}}\right)\, .
%\end{array}
%\end{equation}
Although the transformed $H_{\rm FW}$ is diagonal, it does not have the SUSY structure. The diagonal elements in (\ref{Hdiagonal}) do not form the SUSY partners. However, the eigenstates $|\Psi_\varepsilon^\pm\rangle$ of $H^\pm_{SUSY}$ are also eigenstates of $H^\pm_{\rm FW}$. Hence for $\varepsilon > \varepsilon_0$ we have
\begin{equation}\label{PsiHFWpm}
  H^\pm_{\rm FW}|\Psi_\varepsilon^\pm\rangle = \pm \sqrt{2mc^2\varepsilon}|\Psi_\varepsilon^\pm\rangle\, .
\end{equation}
In addition in the case of unbroken SUSY $\varepsilon_0=M_0^2/2mc^2$ belongs to the spectrum of either $H^+_{\rm SUSY}$ or $H^-_{\rm SUSY}$ or to both and we have
\begin{equation}\label{HFWPsi0}
  \begin{array}{l}
    \displaystyle  |\Psi_{\varepsilon_0}^+\rangle\in{\cal H}^+ \Rightarrow H^+_{\rm FW}|\Psi_{\varepsilon_0}^+\rangle =M_0|\Psi_{\varepsilon_0}^+\rangle\, ,\\
    \displaystyle  |\Psi_{\varepsilon_0}^-\rangle\in{\cal H}^- \Rightarrow H^-_{\rm FW}|\Psi_{\varepsilon_0}^-\rangle =-M_0|\Psi_{\varepsilon_0}^-\rangle\, .
  \end{array}
\end{equation}
Hence the spectrum of $H_{\rm FW}$ and therefore also of $H_D$ is symmetric about the origin with possible exceptions at $\pm M_0$, cf.\ ref.\ [\onlinecite{TH1992}]. The corresponding positive and negative energy eigenstates $|\Psi_{E}^{\rm pos/neg}\rangle$ of $H_D$ can explicitly be calculated from those of $H_{\rm FW}$ by utilizing the unitary transformation (\ref{UFW}). That is,
\begin{equation}\label{PsiD}
  \begin{array}{l}
  |\Psi_{E}^{\rm pos}\rangle = U_{\rm FW}^\dag|\Psi_{\varepsilon}^{+}\rangle\, ,\quad
  H_D|\Psi_{E}^{\rm pos}\rangle= \sqrt{2mc^2\varepsilon}|\Psi_{E}^{\rm pos}\rangle\, ,\\
  |\Psi_{E}^{\rm neg}\rangle = U_{\rm FW}^\dag|\Psi_{\varepsilon}^{-}\rangle\, ,\quad
  H_D|\Psi_{E}^{\rm neg}\rangle= -\sqrt{2mc^2\varepsilon}|\Psi_{E}^{\rm neg}\rangle\, ,
  \end{array}
\end{equation}
and the Dirac eigenvalues $E$ are given by those of the associated SUSY Hamiltonian, $\varepsilon$, via the simple relation
\begin{equation}\label{EDirac}
  E=\pm\sqrt{2mc^2\varepsilon}\,.
\end{equation}
With the help of the SUSY transformation (\ref{SUSYtrafogen}) the positive and negative energy eigenstates explicitly read
\begin{equation}\label{PsiDexplicit+}
  |\Psi_{\varepsilon}^{\rm pos}\rangle =\frac{1}{\sqrt{2+2\sqrt{\varepsilon_0/\varepsilon}}}
  \left(
  \begin{array}{c}
    \left(1+\sqrt{\varepsilon_0/\varepsilon}\right)|\Psi_{\varepsilon}^{+}\rangle\\
    \sqrt{1-\varepsilon_0/\varepsilon}\,|\Psi_{\varepsilon}^{-}\rangle
  \end{array}
  \right)\, ,
\end{equation}
\begin{equation}\label{PsiDexplicit-}
  |\Psi_{\varepsilon}^{\rm neg}\rangle =\frac{1}{\sqrt{2+2\sqrt{\varepsilon_0/\varepsilon}}}
  \left(
  \begin{array}{c}
    -\sqrt{1-\varepsilon_0/\varepsilon}\,|\Psi_{\varepsilon}^{+}\rangle\\
    \left(1+\sqrt{\varepsilon_0/\varepsilon}\right)|\Psi_{\varepsilon}^{-}\rangle
  \end{array}
  \right)\, .
\end{equation}
In the case of unbroken SUSY when $M_0\in{\rm spec}(H_D)$ the corresponding eigenstates are given by $|\Psi_{\varepsilon_0}^{\rm pos}\rangle=|\Psi_{\varepsilon_0}^{+}\rangle$ and in the case of $-M_0\in{\rm spec}(H_D)$ by $|\Psi_{\varepsilon_0}^{\rm neg}\rangle=|\Psi_{\varepsilon_0}^{-}\rangle$.

In closing this section let us summarize that the spectral properties of a supersymmetric Dirac Hamiltonian (\ref{HD3}) are explicitly given via those of the non-relativistic SUSY Hamiltonian (\ref{HS}).

\section{Path integral representation of the resolvent}
In non-relativistic quantum mechanics, the time-evolution operator can be represented in terms of a Feynman path integral. In fact, the Schr\"odinger Hamiltonian being quadratic in the momentum makes it possible to define the Feynman path integral properly\cite{Feyn1948,Feyn1965}. In the Dirac theory, the Hamiltonian is linear in the momentum. Hence the path integral representation cannot be constructed in a manner analogous to that for the non-relativistic case. Especially, for the SUSY Dirac problem, it is desirable to find an effective SUSY Hamiltonian quadratic in the momentum, corresponding to the Dirac operator. To this end we shall employ the following procedure.

Let us start with the resolvent of the Dirac operator (\ref{HD2})
\begin{equation}\label{G}
  G(z)=\frac{1}{H_D-z},\quad z\in\mathbb{C}\backslash \textrm{spec}(H_D)\,,
\end{equation}
which is an analytical function of $z$ in the complement of the spectrum of $H_D$. Let us write it in the form
\begin{equation}\label{Gg}
  G(z)=(H_D+z)g(z^2)\,.
\end{equation}
Here
\begin{equation}\label{g}
  g(\zeta)=\frac{1}{H_D^2-\zeta}
\end{equation}
is the resolvent of the squared Dirac operator $H_D^2$, defined in the $\zeta$-plane. The iterated resolvent $g(\zeta)$ is easier to handle than the original resolvent $G(z)$ is. While the Dirac operator $H_D$ given in (\ref{HD3}) is not diagonal, the squared operator $H_D^2$ is diagonal as shown in (\ref{HD^2}). Hence the iterated resolvent $g(\zeta)$ can as well be given in the diagonal form,
\begin{equation}\label{g2}
 g(\zeta)=\left(
    \begin{array}{cc}
    g^+(\zeta) & 0 \\
    0 & g^-(\zeta)
  \end{array}
  \right) ,
\end{equation}
with
\begin{equation}\label{gpm}
  g^+(\zeta) = \frac{1}{DD^\dag+M_0^2-\zeta}\, , \quad g^-(\zeta) = \frac{1}{D^\dag D+M_0^2-\zeta}\, .
\end{equation}
The resolvent $G(z)$ of the Dirac operator $H_D$ may be obtained with the help of the diagonal elements of the iterated resolvent, in the form,
\begin{equation}\label{Goperator}
\begin{array}{rcl}
  G(z)&=&\left(H_D + z\right)g(z^2) \\[2mm]
      &=&
  \left(
  \begin{array}{cc}
    (z+M_0)g^+(z^2) & Dg^-(z^2) \\
    D^\dag g^+(z^2) & (z-M_0)g^-(z^2)
  \end{array}
  \right).
\end{array}
\end{equation}

The iterated resolvent $g(\zeta)$ as well as $G(z)$ acts on Hilbert space ${\cal H}$. Hence $|\Psi'\rangle=g(\zeta)|\Psi\rangle\in{\cal H}$ for $|\Psi\rangle\in{\cal H}$, which yields on the coordinate base an integral equation,
\begin{equation}
  \langle\textbf{r}''|\Psi'\rangle=\int_{\mathbb{R}^3}\rmd\textbf{r}'\,\langle\textbf{r}''|g(\zeta)|\textbf{r}'\rangle\langle\textbf{r}'|\Psi\rangle \, ,
\end{equation}
whose kernel $\langle\textbf{r}''|g(\zeta)|\textbf{r}'\rangle$, denote by $g(\textbf{r}'',\textbf{r}';\zeta)$, will be referred to as the iterated resolvent kernel.

Since $H_D^2$ in (\ref{g}) and hence $H_{SUSY}$ defined by (\ref{HS}) are positive semi-definite, the spectrum of the corresponding Schr\"odinger operator, $\textrm{spec}(H_{SUSY})$, is on the non-negative real axis of the $\zeta$-plane. Considering a contour $C$ encircling counter-clockwise all points corresponding to $\textrm{spec}(H_D^2)$ and using Cauchy's integral formula, we can find
\begin{equation}\label{contour}
  \rme^{-\rmi tuH^2_D}=-\frac{1}{2\pi \rmi}\oint_C\rmd\zeta\,\rme^{-\rmi tu\zeta}g(\zeta)\, ,
\end{equation}
where $t$ and $u$ are any real constants. At this point, recall the Schr\"odinger operator $H_{SUSY}$ is defined by (\ref{HS}), we replace $H^2_D$ on the left hand side by $2mc^2H_{SUSY}$. Moreover, we let $u=(2mc^2\hbar)^{-1}$. Then the quantity on the left hand side of (\ref{contour}) can be understood as the unitary evolution operator of the system with the Schr\"odinger Hamiltonian $H_{SUSY}$ if $t$ is identified with the time parameter. In the coordinate representation, (\ref{contour}) may be written as
\begin{equation}\label{contour2}
  K(\textbf{r}'',\textbf{r}';t)=-\frac{1}{2\pi \rmi}\oint_C\rmd\zeta\,\rme^{-\rmi t\zeta /2mc^2\hbar}g(\textbf{r}'',\textbf{r}';\zeta)\, ,
\end{equation}
which relates the resolvent kernel to Feynman's kernel (or the propagator),
\begin{equation}\label{propagator}
  K(\textbf{r}'',\textbf{r}';t)=\langle \textbf{r}''|\rme^{-\rmi tH_{SUSY}/\hbar}|\textbf{r}'\rangle\, .
\end{equation}

We also note that the iterated resolvent can be expressed as
\begin{equation}\label{gint}
  g(\zeta)=\rmi u\int_0^\infty\rmd t \exp\left\{-\rmi t u (H_D^2-\zeta)\right\}\, .
\end{equation}
The integral on the right hand side converges for $\textrm{Im}(u\zeta)>0$ and $\zeta\notin\textrm{spec}(H_D^2)$. Namely it converges on the upper half of the $\zeta$ plane if $u>0$, or on the lower half plane if $u<0$. With our choice $u=(2mc^2\hbar)^{-1}$, we consider the integral defined only for $\textrm{Im}(\zeta)>0$. In the coordinate representation, (\ref{gint}) takes the form
\begin{equation}\label{gcoordiante}
  g(\boldsymbol{r}'',\boldsymbol{r}';\zeta)=\frac{\rmi}{2mc^2\hbar}\int_0^\infty \rmd t\ P_\zeta(\boldsymbol{r}'',\boldsymbol{r}';t)\,,
\end{equation}
where
\begin{equation}\label{Pzeta}
  P_\zeta(\boldsymbol{r}'',\boldsymbol{r}';t)=\langle\boldsymbol{r}''|\exp\left\{-(\rmi t/\hbar)\left(H_{SUSY}-\zeta/2mc^2\right)\right\}|\boldsymbol{r}'\rangle\,,
\end{equation}
which we shall refer to as the promotor\cite{Inomata1986}. The promotor for the Hamiltonian $H_{SUSY}$ is the same in form as the propagator for the effective Hamiltonian $H_{eff}(\zeta)=H_{SUSY}-\zeta/2mc^2$. As the propagator is given in terms of Feynman's path integral, so is the promotor. While the Schr\"odinger Hamiltonian itself is self-adjoint, the effective Hamiltonian is not. However, the Hamiltonian has to be self-adjoint for the time evolution operator, but path integration of the promotor does not require self-adjointness of the effective Hamiltonian.

The corresponding diagonal elements of the iterated resolvent kernel are given by
\begin{equation}\label{gpm2}
\begin{array}{rcl}
g^\pm(\boldsymbol{r}'',\boldsymbol{r}';\zeta) &=&
  \langle\boldsymbol{r}''|g^\pm(\zeta)|\boldsymbol{r}'\rangle\\[2mm]
  &=&\displaystyle\frac{\rmi}{2m c^2\hbar}\int_0^\infty \rmd t\ P^\pm_\zeta(\boldsymbol{r}'',\boldsymbol{r}';t)
\end{array}
\end{equation}
where
\begin{equation}\label{Ppm}
  P^\pm_\zeta(\boldsymbol{r}'',\boldsymbol{r}';t)=\langle\boldsymbol{r}''|\exp\{-(\rmi t/\hbar)H_{eff}^\pm(\zeta)\}|\boldsymbol{r}'\rangle
\end{equation}
and
\begin{equation}\label{Heff}
\begin{array}{rcl}
 H_{eff}^+(\zeta) &=&\displaystyle \frac{1}{2mc^2}\left(D D^\dag +M_0^2-\zeta\right)\, ,\\[2mm]
 H_{eff}^- (\zeta) &=& \displaystyle\frac{1}{2mc^2}\left(D^\dag D  +M_0^2-\zeta\right)\, .
\end{array}\end{equation}

One of the merits of utilizing the iterated resolvent is that the kernel of its components can be represented  by means of Feynman's path integral for effective non-relativistic systems. Let ${\cal L}_\zeta^\pm(\dot{\boldsymbol{r}},\boldsymbol{r},t)$ represent the classical Lagrangian associated with the effective Hamiltonian $H_{eff}^\pm(\zeta)$. Then the promotors can be expressed as Feynman's path integral
\begin{equation}\label{PI}
  P^\pm_\zeta(\boldsymbol{r}'',\boldsymbol{r}';t)=\int_{\boldsymbol{r}(0)=\boldsymbol{r}'}^{\boldsymbol{r}(t)=\boldsymbol{r}''}{\cal D}\boldsymbol{r}
  \exp\left\{\frac{\rmi}{\hbar}\int_0^t\rmd s\ {\cal L}_\zeta^\pm(\dot{\boldsymbol{r}},\boldsymbol{r},s)
  \right\}.
\end{equation}
If the system with the effective non-relativistic Lagrangian is path-integrable, the diagonal elements of the iterated resolvent kernel can be determined via (\ref{gpm2}).

The resolvent kernel $G(\boldsymbol{r}'',\boldsymbol{r}';z)=\langle\boldsymbol{r}''|G(z)|\boldsymbol{r}'\rangle$, or Green's function, of the Dirac operator
is given as the coordinate representation of (\ref{Goperator}); namely
\begin{equation}\label{Gexplicit}
\begin{array}{l}
G(\boldsymbol{r}'',\boldsymbol{r}';z)=\\
~=\displaystyle
  \left(
  \begin{array}{cc}
    (z+M_0)g^+(\boldsymbol{r}'',\boldsymbol{r}';z^2) & D(\boldsymbol{r}'')g^-(\boldsymbol{r}'',\boldsymbol{r}';z^2) \\
    D^\dagger(\boldsymbol{r}'')g^+(\boldsymbol{r}'',\boldsymbol{r}';z^2) & (z-M_0)g^-(\boldsymbol{r}'',\boldsymbol{r}';z^2)
  \end{array}
  \right).
\end{array}
\end{equation}
where $D(\boldsymbol{r}'')$ is the operator in the $\boldsymbol{r}''$ representation, satisfying $\langle\boldsymbol{r}''|D|\boldsymbol{r}'\rangle=D(\boldsymbol{r}'')\delta(\boldsymbol{r}''-\boldsymbol{r}')$. If the two non-zero elements of the iterated resolvent kernel are found in closed form, then all the elements of Green's function must be obtained in closed form.

In obtaining the above results, the SUSY-structure is a sufficient but not a necessary condition. The recent approach is applicable even to the case where the Dirac operator is non-supersymmetric, i.e., not of the form (\ref{HD3}), insofar as its squared Dirac operator is (block) diagonal.
A notable example is the Dirac-Coulomb operator for the Dirac electron in the hydrogen atom problem, which is not in the form (\ref{HD3}) but whose square is diagonalizable. The path integral for the Dirac-Coulomb problem has been explicitly and exactly calculated without SUSY\cite{Kayed1984}.

\section{The free Dirac Hamiltonian}
The simplest example for the general approach discussed above is that of the free Dirac Hamiltonian for which $D^\dag=D=c\boldsymbol{\sigma}\cdot\boldsymbol{p}$ and $M_0=m_0c^2$. Let $m=m_0$ as this choice is physically most natural. For simplicity, we use $m$ for the electronic mass rather than $m_0$ from now on.

First we treat this system by using the path integral representation. The effective Hamiltonian is
\begin{equation}\label{Hfree}
  H_{eff}^\pm(\zeta)=\frac{\boldsymbol{p}^2}{2m}+\frac{mc^2}{2}-\frac{\zeta}{2mc^2}=\frac{\boldsymbol{p}^2}{2m}-\frac{\mu^2(\zeta)}{2m}
\end{equation}
with
\begin{equation}\label{kz}
  \mu^2(\zeta)=\zeta/c^2-m^2c^2,
\end{equation}
The effective Hamiltonian (\ref{Hfree}) is nothing more then that of the Schr\"odinger Hamiltonian for a free non-relativistic particle of mass $m$ with an additive constant acting on ${\cal H}^\pm=L^2(\mathbb{R}^3)\otimes\mathbb{C}^2$. The effective Lagrangian corresponding to (\ref{Hfree}) reads
\begin{equation}\label{Lfree}
  {\cal L}_\zeta^\pm (\dot{\boldsymbol{r}},\boldsymbol{r},t)= \frac{m}{2}\ \dot{\boldsymbol{r}}^2+\frac{\mu^2(\zeta)}{2m}\, .
\end{equation}
As is well known, the path integral for (\ref{Lfree}) can explicitly be calculated \cite{Feyn1948,Feyn1965,Schulman1981}, whose result leads the promotor (\ref{Pzeta}) to the form
\begin{equation}\label{Pfree}
\begin{array}{rl}
P^\pm_\zeta(\boldsymbol{r}'',\boldsymbol{r}';t)=&\displaystyle\left(\frac{m}{2\pi\rmi\hbar t}\right)^{3/2}\times\\[4mm]
  &\displaystyle\exp\left\{
  \frac{\rmi}{\hbar}\left(\frac{m}{2t}(\boldsymbol{r}''-\boldsymbol{r}')^2+\frac{t}{2m}\ \mu^2(\zeta)\right)
  \right\}.
\end{array}
\end{equation}
With the help of the integral formula No.\ 3.478.4 in ref. [\onlinecite{Gradstein}], ${\rm Re}\ a>0$, ${\rm Re}\ b>0$,
\begin{equation}
  \int_0^\infty\rmd t\ t^{-3/2}\exp\left\{-a/t-bt\right\}=\sqrt{\frac{\pi}{a}}\exp\left\{-2\sqrt{ab}\right\},\quad
\end{equation}
and assuming a small positive imaginary part of the mass, i.e., ${\rm Im}(m)>0$, we can reduce the resolvent kernel for the squared Dirac Hamiltonian,
$\boldsymbol{x}=\boldsymbol{r}''-\boldsymbol{r}'$,
\begin{equation}\label{gfree}
g^\pm(\boldsymbol{r}'',\boldsymbol{r}';\zeta)=\frac{1}{4\pi|\boldsymbol{x}|(\hbar c)^2}\exp\{(\rmi/\hbar)\mu(\zeta)|\boldsymbol{x}|\},
\end{equation}
from which follows via (\ref{Gg}) the resolvent kernel for the free Dirac Hamiltonian
\begin{equation}\label{Gfree}
\begin{array}{rl}
G(\boldsymbol{r}'',\boldsymbol{r}';z)=&\displaystyle
  \frac{\rme^{(\rmi/\hbar)\mu(z^2)|\boldsymbol{x}|}}{4\pi|\boldsymbol{x}|(\hbar c)^2}\times\\[4mm]
  &\displaystyle
  \left(
  \rmi\hbar c\frac{\balpha\cdot\boldsymbol{x}}{|\boldsymbol{x}|^2}+c\mu(z^2)\frac{\balpha\cdot\boldsymbol{x}}{|\boldsymbol{x}|}+\beta mc^2 +z
  \right).
\end{array}
\end{equation}
This is the well-known result (see e.g.\ eq.\ (1.263) in [\onlinecite{TH1992}]).

The spectral properties of the free Dirac Hamiltonian can directly be read off from those of the free Schr\"odinger Hamiltonians
\begin{equation}\label{HS+-free}
  H_{SUSY}^\pm=\frac{\boldsymbol{p}^2}{2m}+\frac{mc^2}{2}
\end{equation}
using the plane wave solutions. With $\boldsymbol{k}\in\mathbb{R}^3$ denoting the wave vector the spectrum reads
\begin{equation}\label{phihrfee}
  \varepsilon =\frac{\hbar^2\boldsymbol{k}^2}{2m}+\frac{mc^2}{2}\, .
\end{equation}
The free Dirac spectrum  $E=\pm\sqrt{m^2c^4+c^2\hbar^2\boldsymbol{k}^2}$ is immediately recovered via (\ref{EDirac}). The corresponding positive and negative eigenstates are easily be obtained from the non-relativistic plane-waves solutions via the general relations (\ref{PsiDexplicit+}) and (\ref{PsiDexplicit-}).

\section{The Dirac Oscillator}
As a second less trivial example let us consider the so-called Dirac Oscillator
\cite{MosSzc1989,MorZen1989,Benitez1990,Quesne1991}
characterized by $D =c\bsigma\cdot(\boldsymbol{p}+\rmi m\omega\boldsymbol{r})$ and $M_0=mc^2$. With these definitions the  corresponding Schr\"odinger Hamiltonians (\ref{HS+-}) read
\begin{equation}\label{HSDO}
  H_{SUSY}^\pm=\frac{\boldsymbol{p}^2}{2m}+\frac{m}{2}\omega^2\boldsymbol{r}^2
  \mp\left(\frac{2\omega}{\hbar}\boldsymbol{S}\cdot\boldsymbol{L}+\frac{3}{2}\hbar\omega\right)
  +\frac{mc^2}{2}\, ,
\end{equation}
where
\begin{equation}
  \boldsymbol{L}=\boldsymbol{r}\times \boldsymbol{p}\, ,\quad \boldsymbol{S}=(\hbar/2)\boldsymbol{\sigma}\, ,
\end{equation}
denote the orbital and spin angular momentum operator, respectively. Introducing the Spin-Orbit operator
\begin{equation}\label{K}
  K=2\boldsymbol{S}\cdot\boldsymbol{L}/\hbar^2+1=\boldsymbol{J}^2/\hbar^2-\boldsymbol{L}^2/\hbar^2+1/4\, ,\quad \boldsymbol{J}=\boldsymbol{L}+\boldsymbol{S}\, ,
\end{equation}
which obviously commutes with $H^\pm_{SUSY}$, $\boldsymbol{J}^2$, $J_3$ and $\boldsymbol{L}^2$, the above pair of Hamiltonians reads
\begin{equation}\label{HSDOK}
  H_{SUSY}^\pm=\frac{\boldsymbol{p}^2}{2m}+\frac{m}{2}\omega^2\boldsymbol{r}^2
  \mp\left(K+\frac{1}{2}\right)\hbar\omega
  +\frac{mc^2}{2}\, .
\end{equation}
The ansatz
\begin{equation}\label{PSIDO}
  \Psi^\pm_{njm_j\sigma}(\boldsymbol{r})=R^{(\sigma)}_{nl}(r)\varphi^{(\sigma)}_{jm_j}(\theta,\phi)\,,
\end{equation}
where $\varphi^{(\sigma)}_{jm_j}$ denotes the spin spherical harmonics \cite{BjorkenDrell1964} (see also Appendix A),
will reduce the eigenvalue problem of (\ref{HSDOK}) to that of the standard non-relativistic harmonic oscillator with an additional spin-orbit coupling in three dimensions. The radial wave functions $R^{(\sigma)}_{nl}$ are identified with the well-known radial wave function of the harmonic oscillator \cite{Cardoso2003}.
The eigenvalues of (\ref{HSDOK}) are explicitly given by
\begin{equation}\label{EHO}
  \begin{array}{rcl}
  {\cal E}^+_{n,j,\sigma}&=&\hbar\omega\left[2n+j+1-\sigma(j+1)\right]+mc^2/2\, ,\\
  {\cal E}^-_{n,j,\sigma}&=&\hbar\omega\left[2(n+1)+j+\sigma j\right]+mc^2/2\, ,
  \end{array}
\end{equation}
where $n\in\mathbb{N}_0$ denotes the radial quantum number, the total angular momentum quantum number $j=\frac{1}{2},\frac{3}{2},\ldots$ , takes half odd integers and $\sigma = \pm 1$. Hence, we have $\varepsilon_n={\cal E}^+_{n,j,\sigma} = {\cal E}^-_{n-1,j+1,-\sigma}$.
As explicated in Appendix A let us note that  $l$, $j$ and $\sigma$ are not independent quantum numbers as they are related by $j=l+\sigma/2$. That is for fixed $j$ we have the relation
$R^{(\sigma)}_{nl}=R^{(-\sigma)}_{n,l-\sigma}$. For a detailed discussion of the spectral properties we refer to the work by Quesne \cite{Quesne1991}, here we are only interested in the SUSY ground states.
Obviously we have $\varepsilon_0={\cal E}_{0,j,1}^+=mc^2/2$. That is SUSY is unbroken and the ground state energy, which belongs to the spectrum of $H^+_{SUSY}$ only, is infinitely degenerate as it does not dependent on the total angular quantum number $j$. Indeed, observing that
\begin{equation}\label{DDO}
  D^\dag=c\boldsymbol{\sigma}\cdot\left(\boldsymbol{p}-\rmi m \omega\boldsymbol{r}\right)
   = -\rmi\hbar c\boldsymbol{\sigma}\cdot\boldsymbol{e}_r
     \left(\partial_r+\frac{m\omega}{\hbar}r-\frac{K-1}{r}\right)
\end{equation}
where $\boldsymbol{r}=r\boldsymbol{e}_r$, one realizes that the radial part of the eigenfunctions $\Psi^+_{0,j,m_j,1}$ with $n=0$ and $\sigma=1$,
$$
R^{(+1)}_{0l}(r)\propto r^{l}\exp\left\{-\frac{m\omega}{2\hbar^2}r^2\right\}\, ,\quad l=j-1/2\, ,
$$
are all annihilated by the operator (\ref{DDO}).

Finally let us note that the SUSY transformations (\ref{SUSYtrafogen}) explicitly read
\begin{equation}\label{DOSUSY}
  \begin{array}{rcl}
    D^\dag\Psi^+_{n,j,m_j,1} & = & \rmi\sqrt{2mc^2\hbar\omega2n}\Psi^-_{n-1,j,m_j,-1}  \\[1mm]
    D^\dag\Psi^+_{n,j,m_j,-1} & = & -\rmi\sqrt{2mc^2\hbar\omega2(n+j+1)}\Psi^-_{n,j,m_j,1} \\[1mm]
    D\Psi^-_{n,j,m_j,-1} & = & -\rmi\sqrt{2mc^2\hbar\omega2(n+1)}\Psi^+_{n+1,j,m_j,1}  \\[1mm]
    D\Psi^-_{n,j,m_j,1} & = & \rmi\sqrt{2mc^2\hbar\omega2(n+1+j)}\Psi^+_{n,j,m_j,-1}
  \end{array}
\end{equation}
and can be derived by explicit application of (\ref{DDO}) using the property (\ref{A3}) of appendix A and established recurrence relations of the radial functions of the harmonic oscillator \cite{Cardoso2003}.

\subsection{Path integral representation for the Dirac Oscillator}
The effective Hamiltonians associated with the Dirac oscillator explicitly read
\begin{equation}\label{HeffDO}
  {\cal H}_{eff}^\pm(\zeta) =\frac{\boldsymbol{p}^2}{2m}+\frac{m}{2}\omega^2\boldsymbol{r}^2
  \mp\left(K+\frac{1}{2}\right)\hbar\omega
  -\frac{\mu^2(\zeta)}{2m}\,
\end{equation}
Due to the spherical symmetry of above Hamiltonian the promotor associated with it can be expressed in a partial wave expansion of the form
\begin{equation}\label{PDO}
\begin{array}{rl}
  P^\pm_\zeta(\boldsymbol{r}'',\boldsymbol{r}';t)=&
  \displaystyle
    \sum_{j\sigma}P^\pm_{\zeta,l}(r'',r';t)\times\\
  &\displaystyle  \sum_{m_j=-j}^j\varphi^{(\sigma)}_{jm_j}(\theta'',\phi'')\bar{\varphi}_{jm_j}^{(\sigma)}(\theta',\phi')
\end{array}
\end{equation}
where the radial promotor can be expressed in terms of a radial path integral
\begin{equation}\label{PIDO}
  P^\pm_{\zeta,l}(r'',r';t)=\int_{r(0)=r'}^{r(t)=r''}{\cal D}r
  \exp\left\{\frac{\rmi}{\hbar}\int_0^t\rmd s\ {\cal L}_{\rm eff}^\pm(\dot{r},r,t)
  \right\}
\end{equation}
with effective radial Lagrangian, $\kappa=\sigma (j+1/2)$,
\begin{equation}\label{LDO}
\begin{array}{rl}
{\cal L}_{\rm eff}^\pm(\dot{r},r,t) =&\displaystyle
  \frac{m}{2}\dot{r}^2-\frac{m}{2}\omega^2r^2-\frac{l(l+1)\hbar^2}{2mr^2}+\\[2mm]
  &\displaystyle\frac{\mu^2(\zeta)}{2m}\pm\hbar\omega(\kappa+1/2)
\end{array}
\end{equation}
which in essence is that of the radial harmonic oscillator in three dimension and its path integration has explicitly been calculated by Peak and Inomata \cite{Peak1969,InomataJunker94} resulting in
\begin{widetext}
\begin{equation}\label{PlDO}
  P^\pm_{\zeta,\ell}(r'',r';t)=\displaystyle
    \frac{m\omega(r''r')^{1/2}}{\rmi\hbar\sin \omega t}
  \displaystyle\exp\left\{
      \frac{\rmi m\omega}{2\hbar}(r''^2+r'^2)\cot\omega t +\frac{\rmi t\mu^2_\pm(\zeta)}{2\hbar m}
    \right\}\,{\rm I}_{\ell+1/2}\left(\case{m\omega r'r''}{\rmi\hbar\sin\omega t}\right)
\end{equation}
with $\mu_\pm^2(\zeta)=\mu^2(\zeta)\pm2m\hbar\omega(\kappa+1/2)$ and ${\rm I}_{\ell +1/2}$ denotes the modified Bessel function.
The $t$-integration can be performed using the integral formula (derived from 6.669.4 in [\onlinecite{Gradstein}])
\begin{equation}\label{t-integral}
\displaystyle
  \int_0^\infty \rmd q\frac{\rme^{2 \nu q}}{\sinh q}\exp\left\{-\case{1}{2}(a+b)t\coth q\right\}{\rm I}_{2\rho}\left(\case{t\sqrt{ab}}{\sinh q}\right)
\displaystyle
  = \frac{\Gamma(1/2+\mu-\nu)}{t\sqrt{ab}\Gamma(1+2\mu)}\,{\rm W}_{\nu,\rho}(at){\rm M}_{\nu,\rho}(bt)
\end{equation}
with integrability conditions $a>b$, ${\rm Re}(1/2+\rho-\nu)>0$ and ${\rm Re}\ t>0$.
The explicit integration yields
\begin{equation}\label{glDO}
\displaystyle
\int_0^\infty \rmd t\ P^\pm_{\zeta,\ell}(r'',r',t)=
  \frac{(r''r')^{-1/2}}{\rmi\omega}\frac{\Gamma(\rho-\nu^\pm+1/2)}{\Gamma(1+2\rho)}
  {\rm W}_{\nu^\pm ,\rho}(r_>^2 m\omega/\hbar){\rm M}_{\nu^\pm ,\rho}(r_<^2 m\omega/\hbar)
\end{equation}
where $\rho=\ell/2+1/4$, $\nu^\pm(\zeta)=\mu_\pm^2(\zeta)/4m\omega\hbar$, $r_>=\max\{r'',r'\}$, $r_<=\min\{r'',r'\}$ and ${\rm W}_{\rho,\nu}$ and ${\rm M}_{\rho,\nu}$ are the Whittaker functions. We finally arrive at the iterated Green's function for the Dirac oscillator
\begin{equation}\label{gDO}
  g^\pm(\boldsymbol{r}'',\boldsymbol{r}';\zeta)=\displaystyle
    \sum_{j\sigma}g^\pm_{\ell}(r'',r';\zeta)
    \displaystyle\sum_{m_j=-j}^j\varphi_{jm_j}^{(\sigma)}(\theta'',\phi'')\bar{\varphi}_{jm_j}^{(\sigma)}(\theta',\phi')
\end{equation}
with
\begin{equation}\label{glDO2}
g^\pm_{\ell}(r'',r';\zeta)=\displaystyle\frac{\Gamma(\rho-\nu^\pm(\zeta)+1/2)}{2mc^2\hbar\omega(r''r')^{1/2}\Gamma(1+2\rho)}
  {\rm W}_{\nu^\pm,\ \rho}(r_>^2 m\omega/\hbar){\rm M}_{\nu^\pm,\ \rho}(r_<^2 m\omega/\hbar)\,,
\end{equation}
\end{widetext}
and $\bar{\varphi}$ denoting the complex conjugated and transposed spinor of $\varphi$.
From the poles of the Green's function in the complex $z$-plane we may deduce the spectrum. These poles occur in the Gamma function of the numerator of the iterated radial Green's function (\ref{glDO2}), that is, for
\begin{equation}\label{DOpoles}
  \nu^\pm_n(\zeta)=n+\rho+1/2,\quad n\in \mathbb{N}_0\,.
\end{equation}
Hence the poles occur at
\begin{equation}\label{DOSpec}
\textstyle
  \zeta^\pm=m^2c^4+2mc^2\hbar\omega\left[2n+j+\case{3}{2}\pm\case{1}{2}\mp\sigma\left(j+\case{1}{2}\right)-\case{\sigma}{2}\right]
\end{equation}
and the spectrum of the Dirac oscillator explicitly reads
\begin{equation}
\begin{array}{l}
  E_{n,j,\sigma}^+ = mc^2\sqrt{1+\frac{2\hbar\omega}{mc^2}\left[2n+(j+1)-\sigma(j+1)\right]}\, ,\\[2mm]
  E_{n,j,\sigma}^- = -mc^2\sqrt{1+\frac{2\hbar\omega}{mc^2}\left[2(n+1)+j+\sigma j\right]}\,.
\end{array}
\end{equation}
Obviously we have $E^-_{n-1,j+1,-\sigma}=-E^+_{n,j,\sigma}$, which implies that the spectrum is symmetric about the origin with the exception of the ground state energy $E^+_{0,j,1}=mc^2$ which belongs only to the positive part of the spectrum. Note that the spectrum can also be obtained directly from (\ref{EHO}) using the established relation (\ref{EDirac}).

\section{A generalisation of the Dirac Oscillator}
Let us now consider a generalisation of the Dirac oscillator characterised by
\begin{equation}\label{gDOU}
D=c\boldsymbol{\sigma}\cdot\left(\boldsymbol{p}+\rmi\hbar\boldsymbol{\nabla}U\right)\, , \quad M_0=mc^2\,.
\end{equation}
Here the "superpotential" $U$ is initially assumed to be an arbitrary function on $\mathbb{R}^3$. The corresponding Schr\"odinger Hamiltonians now read
\begin{equation}\label{HSU}
  H_{SUSY}^\pm=\frac{\boldsymbol{p}^2}{2m}+\frac{\hbar^2}{2m}\left( \left(\boldsymbol{\nabla}U\right)^2\mp\Delta U\right)
          \mp\frac{2}{m}\boldsymbol{S}\cdot\left(\boldsymbol{\nabla}U\times\boldsymbol{p}\right)\,.
\end{equation}
Obviously SUSY is unbroken if one of the functions
\begin{equation}\label{Psi0U}
  \Psi^\pm_{0,\lambda}(\boldsymbol{r})\propto\exp\{\mp U(\boldsymbol{r})\}\chi_\lambda
\end{equation}
belongs to the Hilbert space ${\cal H}^\pm$, respectively. In the above $\chi_\lambda$ is an arbitrary constant 2-spinor. Note, that both $\Psi^+_{0,\lambda}$ and $\Psi^-_{0,\lambda}$ cannot be simultaneous ground states as, for example, a rapidly increasing superpotential $U(\boldsymbol{r})\to \infty$ for $\boldsymbol{r}\to \infty$ will lead to an normalizable $\Psi^+_{0,\lambda}$ but then $\Psi^-_{0,\lambda}$ is not square integrable.

In order to be a bid more explicit let us consider that the superpotential is a spherical symmetric function. In that case $\boldsymbol{\nabla}U=U'(r)\boldsymbol{e}_r$ and thus $U'$ describes a so-called tensor potential \cite{Furnstahl1998}. The quadratic superpotential $U(r) = (m\omega/2\hbar)r^2$ gives rise to a linear tensor potential and characterises the previously discussed Dirac Oscillator. For a recent discussion on generalisation of the Dirac oscillator see, for example, \cite{Akcay2009,Zarrin2010}. For a general spherical symmetric $U$ the corresponding Schr\"odinger Hamiltonians take the form
\begin{equation}\label{gU}
  H^\pm_{SUSY} =\frac{\boldsymbol{p}^2}{2m}+\frac{\hbar^2}{2m}
  \left(U'\,^2(r)\mp U''(r)\right)\mp\frac{\hbar^2U'(r)}{mr}K.
\end{equation}
The adjoint of above Dirac operator (\ref{gDOU}) reduces to
\begin{equation}\label{gDDO}
  D^\dag = -\rmi\hbar c\boldsymbol{\sigma}\cdot\boldsymbol{e}_r
     \left(\partial_r-\frac{K-1}{r}+U'(r)\right)
\end{equation}
and the condition for unbroken SUSY, that is $D^\dag\Psi^+_{0,\lambda}=0$ with $\Psi^+_{0,\lambda}=R^{(+1)}_{0l}\varphi^{(+1)}_{jm_j}$ leads to a radial wave function of the form
$$
R^{(+1)}_{0l}\propto r^{l}\exp\left\{-U(r)\right\} ,\quad l=j-1/2\, ,
$$
which becomes normalizable if the potential $U$ is well-behaved at the origin and increases rapidly enough for large $r$.
Note that for $\sigma = -1$ there exists no normalizable SUSY ground state. Let us also mention that for a rapidly decreasing superpotential, $U(r)\to -\infty$ as $r\to \infty$, SUSY is also unbroken with ground state $\Psi^-_{0,\lambda}=R^{(-1)}_{0l}\varphi^{(-1)}_{jm_j}$ belonging to the sector $\sigma = -1$ and
$$
R^{(-1)}_{0l}\propto r^{l+1}\exp\left\{U(r)\right\} ,\quad l=j+1/2\, .
$$
In either case the SUSY ground state is infinitely degenerate as in the special case of the Dirac oscillator.
%%%%%%%%%%%%%%%%%%%%%%%%%%%%%%%%%%%%%%
\subsection{The case of a linear superpotential}
Let us consider as a simple non-trivial case that of a linear superpotential
\begin{equation}\label{linU}
  U(r) = \gamma r\, , \quad \gamma >0\, .
\end{equation}
which leads to the following pair of Schr\"odinger Hamiltonians
\begin{equation}\label{gDOH}
  { H}^\pm _{SUSY} =\frac{\boldsymbol{p}^2}{2m}+\frac{\hbar^2}{2m}
  \left(\gamma^2 \mp\frac{2\gamma}{r}K\right)\, .
\end{equation}
As the eigenvalues $\kappa=\sigma(j+1/2)$ of the spin-orbit operator $K$ are non-zero integers, the above Hamiltonian ${ H}_{SUSY}^+$ characterizes a quantum particle under the influence of a Coulomb-like spin-orbit interaction which is attractive for the positive eigenvalues of $K$ where spin and orbital angular momentum are aligned, $\sigma = +1$, and repulsive for the case $\sigma = -1$. The situation for ${H}_{SUSY}^-$ is just the opposite.

Let's first consider the subspace $\sigma = +1$ and make the following ansatz for the eigenvalue problem
\begin{equation}\label{linplus}
  {H}_{SUSY}^\pm R_{nl}^{(+)}\varphi^{(+)}_{jm_j}={\cal E}^\pm_{n,j,+1} R_{nl}^{(+)}\varphi^{(+)}_{jm_j}\,.
\end{equation}
As in this subspace the spin-orbit operator takes the positive eigenvalues $\ell + 1$ this eigenvalue problem is reduced to that of the standard non-relativistic Coulomb problem
\begin{equation}\label{Coul+}
\begin{array}{r}
\displaystyle
  \left(
  \frac{\boldsymbol{p}^2}{2m}\mp\frac{\hbar^2\gamma(l +1)}{mr}+\frac{\hbar^2\gamma^2}{2m}
  \right)
  R_{nl}^{(+)}\varphi^{(+)}_{jm_j}=~~~~~\\
{\cal E}^\pm_{n,j,+1} R_{nl}^{(+)}\varphi^{(+)}_{jm_j}
\end{array}
\end{equation}
with effective charge $e^2=\hbar^2\gamma(l + 1)/m$. Hence the eigenvalue problem is immediately solved when the $R^{(\sigma)}_{nl}$ are taken to be the radial eigenfunction for the Coulomb problem. With the well-known eigenvalues of the non-relativistic Coulomb problem $-(me^4/2\hbar^2)(n+l+1)^{-2}$ the discrete eigenvalues are obtained
\begin{equation}\label{Edisc+}
{\cal E}^+_{n,j,+1}=\frac{\hbar^2\gamma^2}{2m}\left(1-\left(\frac{\ell + 1 }{n+\ell+1}\right)^2\right)\, ,\quad n\in\mathbb{N}_0.
\end{equation}
The corresponding continuous spectrum of ${ H}_{SUSY}^+$ coincides with the full spectrum of ${H}_{SUSY}^-$ and may be written as
\begin{equation}\label{Econt}
  {\cal E}^\pm_{\boldsymbol{k},+1}=\frac{1}{2m}\left(\hbar^2\boldsymbol{k}^2+\hbar^2\gamma^2\right)\, , \boldsymbol{k}\in\mathbb{R}^3\, .
\end{equation}

In a similar manner one finds the discrete spectrum for the $\sigma = -1$ subspace where $l=j+1/2\neq 0$
\begin{equation}\label{Edisc-}
{\cal E}^-_{n,j,-1}=\frac{\hbar^2\gamma^2}{2m}\left(1-\left(\frac{\ell}{n+1+\ell}\right)^2\right)\, .
\end{equation}
Obviously we have recovered again the relation $\varepsilon={\cal E}^+_{n,j,+1} = {\cal E}^-_{n+1,j,-1}$ as expected. This finally leads to the discrete eigenvalues of the Dirac operator as follows
\begin{equation}\label{Elin+}
\textstyle
  E_{n,j,+1}^+=mc^2\left[1+\frac{\hbar^2\gamma^2}{m^2c^2}\left(1-\left(\frac{j+1/2}{n+j+1/2}\right)^2\right)\right]^{1/2}\, ,
\end{equation}
\begin{equation}\label{Elin-}
\textstyle
  E_{n,j,-1}^-=-mc^2\left[1+\frac{\hbar^2\gamma^2}{m^2c^2}\left(1-\left(\frac{j+1/2}{n+j+3/2}\right)^2\right)\right]^{1/2} \, .
\end{equation}
The continuous spectrum is given by
\begin{equation}\label{Elinc}
  E^\pm_{\boldsymbol{k}}=\pm mc^2\left[1+\frac{\hbar^2\gamma^2}{m^2c^2} + \frac{\hbar^2\boldsymbol{k}^2}{m^2c^2}\right]^{1/2}\,.
\end{equation}
In addition the positive and negative Dirac eigenstates for the linear superpotential directly follow from those of the non-relativistic Coulomb problem via the general relations established in Section II.

Finally, let us note that a logarithmic superpotential of the form $U(r)=\gamma\ln(r/r_0)$ also leads to an exactly solvable SUSY Hamiltonian.

\subsection{Path integral representation for the case of a linear superpotential}
Similar to the previous discussion on the Dirac oscillator the promotor can be expressed in terms of partial waves (cf.\ eq.\ (\ref{PDO}) and (\ref{PI})) with effective radial Lagrangian
\begin{equation}\label{Llin}
  {\cal L}_{\rm eff}^\pm(\dot{r},r)
  =\frac{m}{2}\dot{r}^2-\frac{l(l+1)\hbar^2}{2mr^2}-\frac{\hbar^2}{2m}\left(\gamma^2\pm\frac{2\gamma\kappa}{r}\right)+\frac{\mu^2(\zeta)}{2m}
\end{equation}
For the $1/r$-potential the path integral cannot directly be evaluated. However, with the help of the so-called local space time transformation \cite{Inomata1986,Junker1990,Inomata1992} it may be reduced to that of a harmonic potential. To be more explicit, let's introduce a new radial variable $s=\sqrt{r}$ and a new time $\tau$ with $\rmd t = 4s^2\rmd\tau$ the integral expression (\ref{gpm2}) for the radial Green's function (\ref{gDO}) can be put into the form
\begin{equation}\label{gLin}
\begin{array}{rl}
  g^\pm_\ell(r',r',\zeta)
  &\displaystyle
  =\frac{\rmi}{2mc^2\hbar}\int_0^\infty\rmd t P_{\zeta\ell}^\pm(r'',r',t)\\[4mm]
  &\displaystyle
  =\frac{\rmi\sqrt{r''r'}}{mc^2\hbar}\int_0^\infty\rmd \tau \tilde{P}_{\zeta\ell}^\pm(s'',s',\tau)
\end{array}\end{equation}
where
\begin{equation}\label{PIlin2}
  \tilde{P}_{\zeta\ell}^\pm(s'',s',\tau)=\int_{s'=s(0)}^{s''=s(\tau)}{\cal D}s
  \exp\left\{\frac{\rmi}{\hbar}\int_0^\tau\rmd\tau\tilde{\cal L}^\pm_{eff}\right\}
\end{equation}
now represents a path integral for a harmonic oscillator in the new space-time co-ordinates as the new effective Lagrangian reads
\begin{equation}\label{Llin2}
  \tilde{\cal L}^\pm_{eff} =
  \frac{m}{2}\left(\frac{\rmd s}{\rmd \tau}\right)^2 -\frac{\Lambda(\Lambda+1)\hbar^2}{2ms^2} -\frac{m}{2}\Omega^2s^2 \pm\frac{4\hbar^2\gamma\kappa}{m}
\end{equation}
where $\Lambda = 2\ell+1/2$ and $\Omega=(2/m)\sqrt{\hbar^2\gamma^2-\mu^2(\zeta)}$. Hence we immediately obtain the result
\begin{equation}\label{Pl}
\begin{array}{rl}
  \tilde{P}^\pm_{\zeta,\ell}(s'',s',\tau)=
    &\displaystyle
    \frac{m\Omega(s''s')^{1/2}}{\rmi\hbar\sin \Omega \tau}{\rm I}_{\Lambda+1/2}\left(\case{m\Omega s's''}{\rmi\hbar\sin\Omega \tau}\right)\times\\
    &\displaystyle
    \exp\left\{
      \frac{\rmi m\Omega}{2\hbar}(s''^2+s'^2)\cot\Omega t \pm \frac{4\rmi \hbar \gamma \kappa}{m}\tau
    \right\}
\end{array}
\end{equation}
which obviously is identical in form with that of the Dirac oscillator discussed in section V subsection A. Hence we can directly perform the $\tau$-integration using formula (\ref{t-integral}) resulting in
\begin{equation}\label{glDO3}
\begin{array}{rl}
g^\pm_{\ell}(r'',r',\zeta)=&\displaystyle
  \frac{(r''r')^{1/4}}{mc^2\hbar\Omega}\frac{\Gamma(\rho-\nu^\pm+1/2)}{\Gamma(1+2\rho)}\times\\[4mm]
  &{\rm W}_{\nu^\pm,\rho}(r_> m\Omega/\hbar){\rm M}_{\nu^\pm,\rho}(r_< m\Omega/\hbar).
\end{array}
\end{equation}
where $\rho=\Lambda/2+1/4=\ell+1/2$ and $\nu^\pm=\pm 2\hbar\gamma\kappa/\Omega m$. Again the poles of $g^\pm_\ell$ will lead to the eigenvalues of the associated Dirac Hamiltonian and occur at the poles of the Gamma function in the numerator, that is, at $\nu^\pm = n+\ell+1$ with $n\in\mathbb{N}_0$. For the positive energy sector they can only occur for positive $\kappa=\ell+1$ that is $\sigma = 1$ and with $\nu^+=2\hbar\gamma(\ell+1)/\Omega m$ the final result reads
\begin{equation}
  E^+_{n,\ell , 1}=\sqrt{m^2c^4+\hbar^2c^2\gamma^2\left(1+\left(\frac{\ell+1}{n+\ell+1}\right)^2\right)}.
\end{equation}
For the negative energy sector the occurrence of poles requires that $\kappa = -\ell < 0$, $\sigma =-1$ and leads to
\begin{equation}
  E^-_{n,\ell ,-1}=-\sqrt{m^2c^4+\hbar^2c^2\gamma^2\left(1+\left(\frac{\ell}{n+\ell+1}\right)^2\right)}.
\end{equation}
Finally let us note that the Green's function also leads to the continuous spectrum of the system. In fact, the branch cut occurring in the definition of $\Omega$ when taken along the negative half line can be parameterized by $\hbar^2\gamma^2-\mu^2(\zeta)=-\lambda^2$ with $\lambda\in\mathbb{R}_+$ and directly leads to
\begin{equation}
  E^\pm_{\boldsymbol{k}}=\pm\sqrt{m^2c^4+\hbar^2\gamma^2c^2+\hbar^2\boldsymbol{k}^2c^2}\, ,
\end{equation}
where we have identified $\lambda = \hbar|\boldsymbol{k}|$. These eigenvalues are indeed identical to those given in (\ref{Elin+}), (\ref{Elin-}) and (\ref{Elinc}) by noting the relation $j=l+\sigma/2$.

It is worth remarking that a logarithmic superpotential of the form $U(r)=\gamma\ln(r/r_0)$ as mentioned above the path integration can be done in closed form, too.

\section{Conclusions}
In this paper we have studied the resolvent kernel, or Green's function, of supersymmetric Dirac Hamiltonians. It has been shown that the iterated resolvent $g$ of the squared Dirac Hamiltonian allows for an explicit path-integral representation of Feynman's type. This is due to the fact that squared Dirac Hamiltonian becomes block diagonal and each block is represented by an effective Hamiltonian which is of the form of a non-relativistic Schr\"odinger Hamiltonian. The final resolvent kernel $G$ is then simply obtained by differentiation, cf.\ eq.\ (\ref{Gexplicit}). We have explicitly treated the free particle case leading to a path integral of the non-relativistic free particle; the Dirac oscillator whose path integral representation was that of a non-relativistic harmonic oscillator with an additional spin-orbit coupling; and a generalisation of the Dirac oscillator where we have studied the special case of a linear superpotential leading us to a non-relativistic path integral of the Coulomb type, which can explicitly be calculated. The path integral representation may also be useful in obtaining semi-classical approximations of a supersymmetric Dirac Hamiltonian, for example, in case were the path integral is not exactly solvable.

We have also shown in this paper that once the spectral properties, i.e.\ the eigenvalues and eigenfunctions, of the associated non-relativistic SUSY partner Hamiltonians $H^\pm_{SUSY}$ are explicitly known, the spectral properties of the Dirac Hamiltonian immediately follow. The eigenvalues $E$ of the Dirac Hamiltonian are explicitly given in terms of the eigenvalues $\varepsilon$ of the non-relativistic $H^\pm_{SUSY}$ via the simple relation (\ref{EDirac}). The corresponding eigenstates are given by those of $H^\pm_{SUSY}$ via the relation (\ref{PsiDexplicit+}) and (\ref{PsiDexplicit-}).

In the main text we have only considered systems with unbroken SUSY. However, the general approach is valid for good and broken SUSY. In Appendix B we briefly realise the free Dirac Hamiltonian in the so-called supersymmetric representation of the Dirac matrices leading to a broken SUSY as discussed in the general approach. The resulting partner Hamiltonians (\ref{HSbroken}) are identical to those of the unbroken realisation (\ref{HS+-free}) discussed in the main text.

Our examples were all based on a Dirac Hamiltonian in three dimensions but the discussion of sections II and III are also applicable to Dirac Hamiltonians in one and two dimensions in an obvious way. Appendix C briefly discusses the one-dimensional Dirac oscillator and its generalization, the relativistic Witten model. Further examples of supersymmetric Dirac Hamiltonians are briefly discussed in the textbook by Thaller \cite{TH1992}. In particular the Dirac particle in an external magnetic field, cf.\ eq.\ (\ref{HD}), exhibits SUSY and its corresponding partner Hamiltonians are given by the usual Pauli-Hamiltonian. Hence, whenever that Pauli problem is solvable, as for example for a constant magnetic field, the corresponding Dirac problem is also solved.

Finally let us mention that our approach also applies to radial Dirac Hamiltonians in flat space as well as in central back grounds like de Sitter and anti-de Sitter space \cite{Cotaescu99}. It might also be applied to cases were no exact SUSY structure and Foldy-Wouthuysen transformation exists using proper non-relativistic approximations as recently discussed by Jentschura and Noble \cite{JN14}.
%%%%%%%%%%%%%%%%%%%%%%%%%%%%%%%%%%%%%%%%
\appendix
%%%%%%%%%%%%%%%%%%%%%%%%%%%%%%%%%%%%%%%%
\section{Spin Spherical Harmonics}
In this appendix we briefly review the basic properties of the spin spherical harmonics following Bjorken and Drell \cite{BjorkenDrell1964} (see also the textbook by Thaller \cite{TH1992})
\begin{equation}\label{RSH}
  \varphi^{(\sigma )}_{jm_j}(\theta,\phi) = \left(
    \begin{array}{c}
      \sqrt{\frac{l+1/2+\sigma m_j}{2l+1}}\,Y_l^{m_j-1/2}(\theta,\phi) \\
      \sigma \sqrt{\frac{l+1/2-\sigma m_j}{2l+1}}\,Y_l^{m_j+1/2}(\theta,\phi)
    \end{array}
  \right)\, ,
\end{equation}
where $Y_l^{m_l}$ denotes the usual spherical harmonics defined on the unit sphere $S^2$. In the above definition the total and orbital angular momentum quantum numbers $j$ and $l$ are not independent and obey the relation $j=l+\sigma/2$, where $\sigma = \pm 1$. Note that $l\in\mathbb{N}_0$ and $l=0$ is only allowed for the case $\sigma = +1$. The  $\varphi^{(\sigma)}_{jm_j}$ are simultaneous eigenfunctions of $\boldsymbol{J}^2$, $J_3$, $\boldsymbol{L}^2$, and the spin orbit operator $K$.
\begin{equation}\label{varphi}
  \begin{array}{rcll}
  \boldsymbol{J}^2 \varphi^{(\sigma)}_{jm_j} & = & j(j+1)\hbar^2 \varphi^{(\sigma)}_{jm_j}\, ,\quad & j\in\{\frac{1}{2},\frac{3}{2},\frac{5}{2}\ldots \}\, ,\\[1mm]
  J_3 \varphi^{(\sigma)}_{jm_j} & = &m_j\hbar \varphi^{(\sigma)}_{jm_j} \, ,\quad & -j \leq m_j \leq j\, ,\\[1mm]
  \boldsymbol{L}^2 \varphi^{(\sigma)}_{jm_j} &=& l(l+1)\hbar^2 \varphi^{(\sigma)}_{jm_j}\, ,\quad & j=l+\sigma/2 \, , \\[1mm]
  K \varphi^{(\sigma)}_{jm_j} &= & \kappa\varphi^{(\sigma)}_{jm_j}\,,&\kappa=\sigma(j+1/2)\, .
\end{array}
\end{equation}
They form a complete orthogonal set on the Hilbert space $L^2(S^2)\otimes\mathbb{C}^2$ and thus are most suitable for solving eigenvalue problems of spherical symmetric Hamiltonians like $H_{SUSY}^\pm$ as discussed in the main text. Finally let us mention that they obey the relation \cite{BjorkenDrell1964}
\begin{equation}\label{A3}
  \boldsymbol{\sigma}\cdot\boldsymbol{e}_r \varphi^{(\sigma)}_{jm_j} = \varphi^{(-\sigma)}_{jm_j}\,.
\end{equation}
Also note that in the literature the eigenvalues of the spin-orbit operator are sometimes denoted by $-\kappa$ and the operator is multiplied by $\hbar^2$. For simplicity we prefer a dimensionless $K$ with eigenvalues denoted by $\kappa$. Finally let us mention the obvious relation
\begin{equation}\label{A4}
  \boldsymbol{\sigma}\cdot\boldsymbol{p} = -\rmi\hbar( \boldsymbol{\sigma}\cdot\boldsymbol{e}_r)\left(\partial_r-\frac{K-1}{r}\right) \,,
\end{equation}
which was used several times in the main text.
%%%%%%%%%%%%%%%%%%%%%%%%%%%%%%%%%%%%%%%%
\section{Free Dirac Hamiltonian with broken SUSY}
Despite the fact that in the main text we only considered systems with unbroken SUSY the general approach is clearly valid for unbroken and broken SUSY. As a simple but instructive example with broken SUSY let us consider again the free Dirac Hamiltonian
\begin{equation}\label{Hfreebroken}
  H_D=c\balpha\cdot\boldsymbol{p}-\beta mc^2\,.
\end{equation}
Using the so-called supersymmetric representation of the Dirac matrices \cite{TH1992}
\begin{equation}\label{alpha&betaSUSY}
  \balpha_s=\left(
  \begin{array}{cc}
    0 & \bsigma \\
    \bsigma & 0
  \end{array}
  \right)\, ,\quad
  \beta_s=\left(
  \begin{array}{cc}
    0 & -\rmi \\
    \rmi & 0
  \end{array}
  \right)\, ,
\end{equation}
the free Dirac Hamiltonian takes the explicit form
\begin{equation}\label{HfreeBroken}
  H_D=\left(
  \begin{array}{cc}
    0 & c\bsigma\cdot\boldsymbol{p}-\rmi mc^2 \\
    c\bsigma\cdot\boldsymbol{p}+\rmi mc^2 & 0
  \end{array}
  \right)\, .
\end{equation}
Hence it is a supersymmetric Dirac operator with $D=c\bsigma\cdot\boldsymbol{p}-\rmi mc^2$ and $M_0=0$. In contrast to the discussion in section III here, the supercharge is not self-adjoint but leads to the same SUSY partner Hamiltonians
\begin{equation}\label{HSbroken}
  H^\pm_{SUSY}=\frac{1}{2mc^2}\left(c^2\boldsymbol{p}^2+m^2c^4\right)\, .
\end{equation}
However, here $\varepsilon_0=M^2_0/2mc^2=0$ and hence cannot belong to the spectrum of above SUSY Hamiltonians, that is, SUSY is broken for this choice of supercharge. Nevertheless, $H_{SUSY}^\pm$ share the same plan-wave eigenstates  as in the unbroken SUSY case with same eigenvalues given by (\ref{phihrfee}).
%%%%%%%%%%%%%%%%%%%%%%%%%%%%%%%%%%%%%%%%
\section{One-Dimensional Dirac Operators}
The free Dirac operator in one dimension is represented by  $2\times 2$ Pauli matrices and explicitly reads
\begin{equation}
  H_D = c\sigma_1 p + \sigma_3 mc^2
    = \left(\begin{array}{cc}
              mc^2 & cp \\
              cp & -mc^2
            \end{array}
      \right)
\end{equation}
acting on ${\cal H}=L^2(\mathbb{R})\otimes\mathbb{C}^2$. As in the three-dimensional case we may define a generalized Dirac oscillator via the non-minimal substitution $p\rightarrow p-\rmi \hbar U'(x)\sigma_3$, where $U:\mathbb{R}\mapsto\mathbb{R}$ denotes the superpotential, giving rise to the standard non-relativistic SUSY Hamiltonians of the Witten model shifted by $\varepsilon_0=mc^2/2$,
\begin{equation}\label{HSgen1D}
  H^\pm_{SUSY}=\frac{p^2}{2m}+\frac{\hbar^2}{2m}\left(U'\,^2(x)\mp U''(x)\right)+\frac{mc^2}{2}\, ,
\end{equation}
Hence, whenever the eigenvalue problem of such a non-relativistic SUSY Hamiltonian can be solved one immediately has the solution for the Dirac problem.
Indeed, let $\epsilon_n=\varepsilon_n-mc^2/2$ denote the eigenvalues of the unshifted Witten model $H^\pm_{SUSY}-\varepsilon_0$ we immediately have the Dirac spectrum
\begin{equation}\label{DiracWittenSpec}
  E_n^\pm=\pm mc^2\sqrt{1+\frac{2\epsilon_n}{mc^2}}\,.
\end{equation}
The corresponding eigenstates are found similarly via relation (\ref{PsiD}).
In the non-relativistic limit we recover the spectrum of the Witten model,
\begin{equation}\label{NRLimit}
  \lim_{c\to\infty}\left(E^+_n - mc^2\right)=\epsilon_n\,
\end{equation}
confirming that the one-dimensional Dirac operator
\begin{equation}\label{DiracWitten}
  H_D=\left(
  \begin{array}{cc}
    mc^2& c\left(p+\rmi\hbar U'(x)\right) \\
    c\left(p-\rmi\hbar U'(x)\right) & -mc^2
  \end{array}
    \right)
\end{equation}
is indeed the relativistic version of Witten's non-relativistic SUSY model characterised by the SUSY partner Hamiltonians (\ref{HSgen1D}).

The special case $U(x)=\frac{m\omega}{2\hbar}x^2$ leads to the one-dimensional Dirac oscillator Hamiltonian
\begin{equation}\label{DO1D}
  H_D
    = \left(\begin{array}{cc}
              mc^2 & c(p+\rmi m\omega x)\\
              c(p-\rmi m\omega x) & -mc^2
            \end{array}
      \right)\,.
\end{equation}
which obviously is a supersymmetric Dirac Operator with $D=c(p+\rmi m\omega x)$, $M_0=mc^2$ and the SUSY Hamiltonians read
\begin{equation}\label{HSDO1D}
  H^\pm_{SUSY}=\frac{p^2}{2m}+\frac{m}{2}\omega^2 x^2\mp \frac{\hbar\omega}{2}+\frac{mc^2}{2}\, ,
\end{equation}
coinciding, except of the trivial additive constant $mc^2/2$, with the supersymmetric harmonic oscillator in Witten's non-relativistic model \cite{Junker1996}.
Here SUSY is unbroken as the ground state energy $\varepsilon_0=M^2_0/2mc^2=mc^2/2$ and belongs to the spectrum of $H^+_{SUSY}$. The positive eigenvalues are $\varepsilon_n =\hbar\omega n+\varepsilon_0$, $n=1,2,3,\ldots$, and the corresponding eigenstates are given by $\Psi^+_n(x)=\langle x|n\rangle$ and $\Psi^-_n(x)=\langle x|n-1\rangle$ where $|n\rangle$ represents the standard $n$-th eigenstate of the one-dimensional harmonic oscillator. Finally, the complete spectrum of the one-dimensional Dirac oscillator reads
\begin{equation}\label{DO1DSpec}
  E_0^+ = mc^2\, ,\quad E^{\pm}_n =\pm mc^2\sqrt{1+2n\frac{\hbar\omega}{mc^2}}\, ,\quad n\in \mathbb{N}\,.
\end{equation}
Let us note that the positive part of this spectrum coincides with that of the three-dimensional Dirac operator when $\sigma = 1$ and the negative part coincides with the negative part of the three-dimensional case if $\sigma = -1$ is chosen. For a detailed study of the dynamical behavior for this model we refer to ref.\ [\onlinecite{Toyama1997}], which compares the dynamics generated by above $H_D$ with that generated by the corresponding diagonalized Hamiltonian $H_{\rm FW}$.
%%%%%%%%%%%%%%%%%%%%%%%%%%%%%%%%%%%%%%


\begin{thebibliography}{10}
\bibitem{Nic76}H.\ Nicolai, \JPA {\bf 9}, 1497 (1976).
\bibitem{Wit81}E.\ Witten, {\rm Nucl.\ Phys.\ B} {\bf 188}, 513 (1981).
\bibitem{Junker1996}G.\ Junker, {\it Supersymmetric Methods in Quantum and Statistical Physics}, (Springer, Berlin, 1996).
\bibitem{Jackiw1984}R.\ Jackiw, { Phys.\ Rev.\ D} {\bf 29}, 2375 (1984).
\bibitem{Hughesetal86}R.J.\ Hughes, V.A.\ Kosteleck\'y and M.M.\ Nieto, Phys.\ Rev.\ D {\bf 34}, 1100 (1986).
\bibitem{Cooper1988}F.\ Cooper, A.\ Khare, R.\ Musto and A.\ Wipf, {Ann.\ Phys.\ } {\bf 187}, 1 (1988).
\bibitem{Thaller1988}B.\ Thaller, J.\ Math.\ Phys.\ {\bf 29}, 249 (1988).
\bibitem{Beckers1990}J.\ Beckers and N.\ Debergh, {Phys.\ Rev.\ D} {\bf 42}, 1255 (1990).
\bibitem{Thaller1991}B.\ Thaller, in A.\ Boutet de Monvel, P.\ Dita, G.\ Nenciu and R.\ Purice (eds), {\it Recent Developments in Quantum Mechanics}, Mathematical Physics Studies (A Supplementary Series to Letters in Mathematical Physics), vol 12, (Springer, Dordrecht, 1991) 351.
\bibitem{TH1992}B.\ Thaller, {\it The Dirac Equation}, (Springer, Berlin, 1992).
\bibitem{Groverelal2014}T.\ Grover, D.N. Sheng and A.\ Vishwanath, Science {\bf 344}, 280 (2014).
\bibitem{Sarma2011}S.D.\ Sarma, S.\ Adam, E.H.\ Hwang and E.\ Rossi, Rev.\ Mod.\ Phys.\ {\bf 83} 407 (2011).
\bibitem{Ezawa2006}M.\ Ezawa, Phys.\ Lett.\ A {\bf 372}, 924 (2008).
\bibitem{Feyn1948}R.P.\ Feynman, \RMP {\bf 20}, 367 (1948).
\bibitem{Feyn1965}R.P.\ Feynman and A.R.\ Hibbs, {\it Quantum Mechanics and Path Integrals}, (McGraw-Hill, New York, 1965).
\bibitem{Inomata1986}A.\ Inomata, in M.C.\ Gutzwiller, A.\ Inomata, J.R.\ Klauder and L.\ Streit eds., {\it Path Integrals from meV to MeV}, (World Scientific, Singapore, 1986).
\bibitem{Kayed1984}M.A.\ Kayed and A.\ Inomata, {Phys.\ Rev. Lett.} {\bf 53}, 107 (1984).
\bibitem{Schulman1981}L.S.\ Schulman, {\it Techniques and Applications of Path Integration}, (John Wiley \& Sons, New York, 1981).
\bibitem{Gradstein}I.S.\ Gradsteyn and I.M\ Ryzhik, {\it Table of Integrals, Series, and Products}, Edited by A.\ Jeffrey and D.\ Zwillinger, (Academic Press, New York,2007) 7th ed.
\bibitem{MosSzc1989}M.\ Moshinsky and A.\ Szczepaniak, \JPA {\bf 22}, L817 (1989).
\bibitem{MorZen1989}M.\ Morena and A.\ Zentella, \JPA {\bf 22}, L821 (1989).
\bibitem{Benitez1990}J.\ Ben\'itez, R.P.\ Mart\'inez y Romero, H.N.\ N\'u\~nez-Y\'epez and A.L.\ Salas-Brito, \PRL {\bf 64}, 1643 (1990).
\bibitem{Quesne1991}C.\ Quesne { Int.\ J.\ Mod.\ Phys.\ A} {\bf 6}, 1567 (1991).
\bibitem{BjorkenDrell1964}J.D.\ Bjorken and S.D.\ Drell, {\it Relativisitc Quantum Mechanics}, (McGraw-Hill, New York, 1964).
\bibitem{Cardoso2003}J.L.\ Cardoso and R.\ \'Alvarez-Nodarse, \JPA {\bf 36}, 2055 (2003).
\bibitem{Peak1969}D.\ Peak and A.\ Inomata, \JMP {\bf 10}, 1422 (1969).
\bibitem{InomataJunker94} A.\ Inomata and G.\ Junker, in R.\ Wilson and E.A.\ Tanner eds., { \it Noncompact Lie groups and their physical applications}, (Kluwer, Dordrecht, 1994) 199.
\bibitem{Furnstahl1998}R.F.\ Furnstahl, J.J.\ Rusnal and B.D.\ Serot, Nucl.\ Phys.\ A {\bf 632}, 607 (1998).
\bibitem{Akcay2009}H.\ Akcay, Phys.\ Lett.\ A {\bf 373}, 616 (2009).
\bibitem{Zarrin2010}S.\ Zarrinkamar, A.A.\ Rajabi and H.\ Hassanabadi, Ann.\ Phys.\ {\bf 325}, 2522 (2010).
\bibitem{Junker1990}G.\ Junker, \JPA {\bf 23}, L881 (1990).
\bibitem{Inomata1992}A.\ Inomata, H.\ Kuratsuji and C.C.\ Gerry, {\it Path Integrals and Coherent States of SU(2) and SU(1,1)} (World Scientific, Singapore, 1992).
\bibitem{Cotaescu99}I.I.\ Cot\u aescu, Phys.\ Rev. {\bf D60}, (1998).
\bibitem{JN14}U.D.\ Jentschura and J.H.\ Noble, \JPA {\bf 47}, 045402 (1990).
\bibitem{Toyama1997}F.M.\ Toyama, Y.\ Nogami and F.A.B.\ Coutinho, \JPA {\bf 30}, 2585 (1997).

%\bibitem{Kalka1997}Kalka H and Soff G 1997 {\it Supersymmetrie} (Stuttgart: Teubner)\bibitem{Kayed1984}Kayed M A and Inomata A 1984 \PRL {\bf 53} 107
%\bibitem{Alhaidari2004}Alhaidari A D 2004 Int.\ J.\ Theor.\ Phys.\ {\bf 43} 939

\end{thebibliography}
\end{document}